\documentclass[preprint,amsmath,amssymb,aps]{revtex4-1}
\usepackage{graphicx}
\begin{document}

\title{Poker Cash Game: a Thermodynamic Description}

\author{Marco Alberto Javarone}
\email{marcojavarone@gmail.com}
\affiliation{Dept. of Mathematics and Computer Science. University of Cagliari, Cagliari (Italy)\\ Dept. of Humanities and Social Science. University of Sassari, Sassari (Italy)}

\date{\today}
\begin{abstract}
Poker is one of the most popular card games, whose rational investigation represents also one of the major challenges in several scientific areas, spanning from information theory and artificial intelligence to game theory and statistical physics. 
In principle, several variants of Poker can be identified, although all of them make use of money to make the challenge meaningful and, moreover, can be played in two different formats: tournament and cash game.
An important issue when dealing with Poker is its classification, i.e., as a `skill game' or as gambling.
Nowadays, its classification still represents an open question, having a long list of implications (e.g., legal and healthcare) that vary from country to country.
In this study, we analyze Poker challenges, considering the cash game format, in terms of thermodynamics systems. Notably, we propose a framework to represent a cash game Poker challenge that, although based on a simplified scenario, allows both to obtain useful information for rounders (i.e., Poker players), and to evaluate the role of Poker room in this context. 
Finally, starting from a model based on thermodynamics, we show the evolution of a Poker challenge, making a direct connection with the probability theory underlying its dynamics and finding that, even if we consider these games as `skill games', to take a real profit from Poker is really hard.
\end{abstract}

\maketitle
\section{Introduction}

Today understanding Poker games, from the mental aptitide of their players to the underlying probabilistic structure, represents a great challenge for scientists belonging to several communities as psychologists, computer scientists, physicists, and mathematicians~\cite{bowling01,sire01,javarone01,teofilo01}.
In general, these games can be analyzed considering psychological aspects, information theory approaches and analytical descriptions. Notably, approaches based on sociophysics~\cite{barra01,barra02,galam01,loreto01,javarone02} allow to study the role of human behavior~\cite{suzuki01,javarone01}. On the other hand, information theory and analytical approaches allow to identify both new algorithms~\cite{teofilo01,dahl01} in the context of artificial intelligence~\cite{norvig01}, and universal properties of these games~\cite{sire01}.
An interesting problem, when dealing with Poker, is constituted by its classification, i.e., `skill game' or gambling. This issue has not yet been solved, although the nowadays available related answer has a long list of implications~\cite{kelly01,cabot01}. 
A preliminary attempt to solve this question, by using the framework of statistical mechanics, has been developed in~\cite{javarone01}, where the author analyzed the role of rationality in a simplified scenario, referred to the Poker variant called Texas Hold'em~\cite{sklansky01}.
In general, all variants follow a similar logic: rounders (i.e., Poker players) receive a number of cards, and have to decide if to bet or not, by computing the possible combinations they can set with their cards (called \textit{hand}). 
After evaluating if the received hand is promising or not, each rounder can take part to the pot by placing a bet (money or chips), otherwise she/he folds the \textit{hand}.
Therefore, the use of money makes the challenge meaningful, otherwise none would have a reason to fold her/his hand. 
Poker challenges can follow two different formats, i.e., cash game or tournament. During a tournament, rounders pay, only once, an entry fee: a fraction goes into the prize pool, and the remain part is a fee to play. 
Eventually, top players share the prize pool (usually money).
On the other hand, playing Poker in the cash game format means to use real money during the challenge. In this case, rounders can play until they have money and, although there no entry fees to pay, a fraction of each pot is taxed, i.e., a small `rake' is applied.
In this work, we propose a framework to study the evolution of a Poker challenge, considering the cash game format, by a thermodynamic description. In particular, we aim both to model these dynamics to achieve insights, and to link the resulting thermodynamic description with the probability theory tacitly governing these games.
\section{Mapping Poker to a Thermodynamic System} \label{sec:model}
\begin{figure}[!h]
\centering
\includegraphics[scale=0.6]{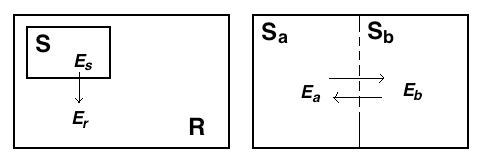}
\caption{Thermodynamic representation of a cash game Poker challenge. On the left, the subsystem $S$ inside the environment $R$, with the arrow indicating the allowed direction of energy transfers, i.e., from $S$ to $R$. On the right, a zoom on the subsystem $S$, showing the two subsystems $S_A$ and $S_B$, representing the two rounders (i.e., $A$ and $B$). Inside $S$, as shown by arrows, the energy can flow from $S_A$ to $S_B$ and vice versa.}
\label{fig:figure_1}
\end{figure}

In this work, we aim to describe cash game Poker challenges by the language of thermodynamics. In particular, since these challenges entail transfers of money among different parts, i.e., rounders and dealers, we assume that the way thermodynamics explains equilibria and energy transfers between systems constitutes a fundamental tool to our investigations.
Firstly, we consider a simple thermodynamic system composed of the subsystem $S$ and its environment $R$.
The total energy of the system $E_T$ is given by the energy of $S$ and that of $R$, i.e., $E_T = E_S + E_R$.
In the proposed model, the environment $R$ is a Poker room, whereas $S$ corresponds to the table where two rounders, say $A$ and $B$, face by a `heads-up' challenge. A `heads-up' is a challenge characterized by the presence of only two rounders. Therefore, we can identify two subsystems of $S$: $S_A$ and $S_B$, corresponding to the two rounders. 
Since Poker challenges are performed following the cash game format, the money is the exchanged quantity, hence mapping money to the energy of the systems come immediate. In doing so, we have $E_A$ and $E_B$ that correspond to the money of $A$ and $B$, respectively.
Therefore $E_S = E_A + E_B$ and, as initial condition, we impose that at $t=0$ rounders have the same amount of money, i.e., $E_A(0) = E_B(0)$.
Figure~\ref{fig:figure_1} offers a pictorial representation of the described system.
During the challenge, some rounds are won by $A$ and others by $B$; hence a fraction of energy is transferred from the subsystem $S_A$ to $S_B$, and vice versa, over time.
The amount of transferred energy $\Delta_S$ corresponds to the total amount of money that flows from $A$ to $B$ and vice versa. For the sake of simplicity, we consider that at each round rounders bet the same amount of money, i.e., pots are constant. In particular, $\Delta_S$ is defined as
\begin{equation}\label{eq:delta_s}
\Delta_S(t) = \Phi_{A,B}(t) + \Phi_{B,A}(t)
\end{equation}
\noindent with $\Phi_{x,y}(t)$ total flow of energy from the subsystem $x$ to $y$, at time $t$. Then, $\Phi_{A,B}$ indicates the total amount of enegy transferred from $S_A$ to $S_B$, as result of all successes of the rounder $B$.
It is worth to note that, in real scenarios, Poker rooms apply a small fee, called `rake', to each pot. Usually, the `rake' corresponds to about $5 \%$ of the pot.
As result, the total energy of $S$ decreases over time by a factor $\Delta_S \cdot \epsilon$, due to energy transfers between $S_A$ and $S_B$.
Since we are dealing with a closed system (i.e., $S$), the loss in energy can be thought in terms of energy reduction due to the entropy's growth $\sigma$. Notably, this concept characterizes the Helmholtz free energy potential $F$
\begin{equation}\label{eq:helmotz}
F = E - T\sigma
\end{equation}
\noindent with $E$ and $T$, energy and temperature of the system, respectively.
In few words, the free energy corresponds to the energy a system can actually use. Then, we can map this concept to our model by the following relation
\begin{equation}\label{eq:free_money}
F_S(t) = E_S(t) -  \Delta_S(t) \cdot \epsilon
\end{equation}
\noindent with $F_S(t)$ free energy of our system, that is available at time $t$, after all energy transfers.
The energy lost by $S$ goes to the environment $R$, hence $E_R = \Delta_S(t) \cdot \epsilon$. It is worth to observe that, as the entropy of a system does, the quantity $\Delta_S(t) \cdot \epsilon$ increases over time, and can never be negative.
\subsection{Evolution of the System}
Now, we focus our attention on the evolution of the system. In particular, we consider one subsystem, i.e., $S_A$ or $S_B$, in order to analyze its amount of energy over time.
Let us consider, for instance, $S_A$ (that represents the rounder $A$) whose evolution can be described by the following relation
\begin{equation}\label{eq:energy_evolution}
E_{A}(t) = E_{A}(0) - \Phi_{A,B}(t) + \Phi_{B,A}(t) \cdot (1 - \epsilon)
\end{equation}
\noindent as the amount of energy in $S_A$ corresponds to the initial amount of energy in this subsystem (i.e., $E_{a}(0)$), minus the amount of energy that flowed to $S_B$ (i.e., $\Phi_{A,B}(t)$), plus the amount of energy that flowed from $S_B$ to $S_A$ (i.e., $\Phi_{B,A}(t)$) reduced of a factor $\epsilon$.
In an equilibrium condition, $\lim_{t \to \infty} E_{A}(t) = E_{A}(0)$, therefore
\begin{equation}\label{eq:equilibrium}
\Phi_{A,B}(t) = \Phi_{B,A}(t) \cdot (1 - \epsilon)
\end{equation}
The primary target of the rounder $A$ is to win all the money of $B$, while avoiding to lose her/his money. Hence, rounder $A$ aims to obtain $E_{A}(t)\ge E_{A}(0)$.
Considering Poker as a skill game~\cite{javarone01}, the flow $\Phi_{x,y}$ depends on the ability of the $y$th rounder. Then, $A$ has a probability $P_a$ to win each round, strongly related to her/his skills.
As discussed before, to simplify the scenario, we suppose that rounders bet always the same amount of money $\frac{\delta}{2}$, forming the pot $\delta$, so that $\Delta_S(t) = \delta \cdot t$.
In doing so, we can define the amount of energy transferred from $S_B$ to $S_A$ as
\begin{equation}\label{eq:energy_flow_a}
\Phi_{B,A} = P_a \cdot \Delta_S
\end{equation}
and the amount of energy transferred from $S_A$ to $S_B$
\begin{equation}\label{eq:energy_flow_b}
\Phi_{A,B} = (1 - P_a) \cdot \Delta_S
\end{equation}
Going back to the equilibrium condition defined in Equation~\ref{eq:equilibrium}, we can write
\begin{equation}\label{eq:equilibrium_probability}
P_a \cdot \Delta_S \cdot (1 - \epsilon) = (1 - P_a) \cdot \Delta_S
\end{equation}
\noindent working a little bit of algebra, from Equation~\ref{eq:equilibrium_probability}, we obtain
\begin{equation}\label{eq:probability_a_epsilon}
P_a = \frac{1}{2 - \epsilon}
\end{equation}
Therefore, we can compute the minimal success probability that the rounder $A$ needs to reach her/his target, i.e., to win (or, at least, to not losing money).
It is worth to highlight that, starting by a thermodynamic description of the system, we can define a relation between the `rake', applied by a Poker room, and the rounders' skills, i.e., their probability to success in Poker cash game.
\subsection{Profits over time}
In light of these results, it is interesting to evaluate both the amount of money rounders can win by playing Poker cash game, and the amount of money the Poker room generates during the challenge.
Considering the rounders's perspective, we are interested in computing the expected value of energy that flows in the subsystem $S_A$, i.e., $<\Delta E_{A}>$ (note that similar considerations hold also for $S_B$). 
The value of $<\Delta E_{A}>$ can be computed as follows
\begin{equation}\label{eq:average_phi}
<\Delta E_{A}> = <\Phi_{B,A}> \cdot (1 - \epsilon) - <\Phi_{A,B}>
\end{equation}
\noindent $<\Phi_{B,A}>$ corresponds to $<\Phi_{B,A}> = P_a \cdot \varphi_{B,A}$, and $<\Phi_{A,B}> = (1- P_a) \cdot \varphi_{A,B}$, with $\varphi_{x,y}$ representing the total energy transfer from $S_x$ to $S_y$, i.e., $\varphi = t \cdot \frac{\delta}{2}$. 
Then, we obtain
\begin{equation}\label{eq:average_phi_rewrite}
<\Delta E_{A}> = P_{A} \cdot  t \cdot \frac{\delta}{2}\cdot (1 - \epsilon) - (1- P_{A}) \cdot  t \cdot \frac{\delta}{2}
\end{equation}
\noindent and we find the following relation
\begin{equation}\label{eq:average_phi_solution}
<\Delta E_{A}> =  t \cdot \frac{\delta}{2} [P_A \cdot (1 - \epsilon) - 1 + P_A] =  t \cdot \frac{\delta}{2} [P_A \cdot (2 - \epsilon) - 1]
\end{equation}
\noindent that is in perfect accordance with results achieved in Equation~\ref{eq:probability_a_epsilon}, as for $P_a = \frac{1}{2 - \epsilon}$ the expected value of energy transferred to $S_A$ is $<\Delta E_{A}> = 0$.
Moreover, it is immediate to note that, considering Equation~\ref{eq:energy_evolution}, the rate of variation of the energy of one subsystem (e.g., $S_x$) corresponds to
\begin{equation}\label{energy_variation}
\frac{dE_x}{dt} = \frac{<\Delta E_x>}{t}
\end{equation}
It is worth to observe that, by all the illustrated equations, it is possible to evaluate the potential gain of a rounder, once her/his winning probability $P_x$ is known.
On the other hand, considering the Poker room perspective, the overall scenario becomes pretty nice, because as we are going to show, its profits can only increases over time without running any risk.
Notably, the environment $R$ (i.e., the Poker room) receives a constant amount of energy, at each time step, equal to $\delta \cdot \epsilon$.
Hence, in the event rounders have the same probability to win (i.e., $P_A = P_B$), it is interesting to compute the number of time steps required to let the Poker room drain almost all their money (i.e., their energy).
Since to perform a round both rounders have to bet the same amount of money, a minimal amount of energy always will remain in the subsystem $S$. In particular, this quantity is equal to $\frac{\delta}{2}$. 
Hence, supposing both subsystems, at time $t$, contain an energy equal to $\frac{\delta}{2}$, the last round entails one subsystem loses completely energy and the other has, at the end, an energy equal to ${\delta}(1-\epsilon)$.
Therefore, the maximum amount of energy that the environment can receive is $2\cdot E_{s}(0) - \delta (1 - \epsilon)$, so that the following relation holds
\begin{equation}\label{eq:poker_room_limit}
\Delta_s \epsilon t = 2 E_{s}(0) - \delta (1 - \epsilon)
\end{equation}
\noindent then, it is possible to compute the number of time steps $t$ to let the Poker room draining almost all the rounders's money:
\begin{equation}\label{eq:poker_room_time}
t = \frac{2 E_{s}(0) - \delta (1 - \epsilon)}{\Delta_s \epsilon}
\end{equation}
Equation~\ref{eq:poker_room_time} shows a direct relation between the time and the rake applied by the Poker room: as the latter increases the time to drain almost all the energy decreases.

\section{Discussion and Conclusions}\label{sec:conclusions}
In this work, we propose a framework for studying Poker challenges in the context of thermodynamics. In particular, we map a simple scenario, where two rounders face, to a thermodynamic system composed of a subsystem $S$ embedded in a larger environment $R$. The former represents the two rounders, whereas the latter the Poker room. 
Remarkably, from a simplified description of the game dynamics, we achieve insights on Poker challenges, in the cash game format.
Even considering this format of Poker as a `skill game' (see \cite{javarone01}), we identify a direct link between the rounders's skills and the fee applied by Poker room, called `rake'.
In doing so, it is possible to know the minimal probability to success a rounder needs to have in order to be a successful player. As shown, gaining by this activity is a very hard task, even for skilled rounders, as they have to keep their probability to win very high.
In real scenarios, many expert rounders are very good and fast in computing winning probabilities for their \textit{hands}, hence they perform online cash game by a `multitabling' strategy: they face at the same time several opponents, with the aim to optimize their profits (obviously, increasing the probability of losing a lot of money).
Moreover, we analyze profits of a Poker room, obtained while rounders play the cash game Poker.
In particular, mapping this profit to the energy of the environment $R$, we achieve the relation $E_R(\epsilon,t) = \delta \epsilon t$, with $\epsilon$ representing the `rake', $\delta$ the pot of each round, and $t$ the number of time steps.
It is worth noting that there are two different situations that allow the Poker room to increase its profits:
\begin{enumerate}
\item Increasing $\epsilon$
\item Increasing $t$
\end{enumerate}
While the first should be kept low (e.g., $\% 5$ or less) as a strategy marketing to attract rounders in the Poker room, the second requires more attention as, in principle, it can lead to a fraudulent strategy, now we briefly illustrate. 
People usually are not worried about frauds in Poker, as they play against other people, and not against a dealer (as in games like the roulette). 
Therefore, in principle, there are no reasons for the electronic dealer to favor a particular rounder in the process of cards distribution. 
Anyway, it is important to highlight that the Poker room does not take an advantage when rounders perform `all-in' actions (i.e., the bet all their money in only one \textit{hand}). 
Then, supposing rounders are rational, i.e., their actions are performed by considering their probability to win each round, a pseudo-random algorithm for cards distribution could be properly defined for generating uncertain scenarios. 
Here, for uncertain scenarios we indicate those situations where both rounders have low winning probabilities, by considering only the information they have (i.e., their \textit{hand} and, in case, common cards).
Therefore, a fraudulent strategy could be implemented by using an algorithm to provide rounders with low winning probability at each hand, in order to avoid they perform `all-in'. 
It is evident that by this strategy, it would be possible to indirectly increasing $t$ for each challenge.
Moreover, it would be also very difficult to find this kind of fraud by analyzing the algorithm, used by a Poker room, if this fraudulent scenario is not considered.
In order to conclude, we would like to emphasize that some of the considerations about the probability to win a cash game challenge can be applied also in the context of financial trading. In particular, for the strategy adopted by `scalpers', i.e., traders that in few seconds open and close a position (i.e., buy and sell financial products as stocks, bonds, etc.), as also in those cases for each transaction the banking system applies a kind of `rake'.

\end{document}